# Sensitivity of External Magnetic Field on the Change in Cross-section of a Toroidal Current

Suman Aich,[1,2,*] Joydeep Ghosh,[1,2] Rakesh L. Tanna,[1] D. Raju,[1,2] Sameer Kumar,[1] and Aditya-U team[1]

[1]Institute for Plasma Research, Bhat, Gandhinagar, Gujarat, India, 382 428
[2]Homi Bhabha National Institute, Training School Complex, Anushaktinagar, Mumbai, India, 400 0944

**ABSTRACT**. Due to any toroidal current column, the magnetic field is found to be sensitive as well as insensitive to its cross-sectional area depending on location of subject point, as predicted by numerical approaches [S. Aich, J. Thakkar, and J. Ghosh, *Plasma Fusion Res.* **17**, 2403055 (2022)], and hence the presence of an angle of invariance is found to be present for any toroidal geometry. Present study aims to validate those numerical observations using the measured magnetic field due to Aditya Upgrade tokamak plasma.

## I. INTRODUCTION

The magnetic field topology plays a crucial role in the confinement and diagnosis of toroidal plasma currents in fusion machines, which are among the most potential candidates amongst non-fuel future energy resources [1-3]. The geometry of such a toroidal plasma current channel with Mega Ampere (MA) order of magnitude and, having a cross-sectional area of tens of centimeters creates an external magnetic field, which neither can be treated as a filamentary current nor like a surface current while considering the external magnetic field (***B***) at the close vicinity of the current column. Though a complete topology of ***B*** can be achieved using well-known magnetic equilibrium fitting (EFIT) [4], the dependency of ***B*** on the cross-sectional area of such current channel is reported in this paper for the first time. Inspite of a minute change in the magnitude of ***B*** due to change in the cross-sectional at a finite distance away from the torus, the modifications in ***B*** are very much crucial for the magnetic diagnostics that are installed at the very vicinity of the toroidal current, carrying MA order of current. Few recent works have addressed this study explicitly and a deep insight about the topology of ***B*** in the vicinity of the current channel have been explored by means of numerical method-based techniques [5-6]. Based on those numerical outcomes, experiments are designed, executed and data are analysed for a successful validation of the consequences.

Present study starts by providing an overview about the major observations, which are established using POISSON SUPERFISH [7] in previous papers [5-6]. Some of those results are re-produced using a newly developed numerical approach in section II and III. Next half of the paper is aimed for experimentation, data analysis and validation and it starts from section IV, which addresses the required plasma diagnostics. Then, a technique for the experimental validation of the numerical outcomes is introduced in section V. Section VI rigorously discusses about the experimental observations and their inferences in validating the numerical predictions. Finally, the entire paper is concluded by providing a precise summary of this work in section VII.

## II. NUMERICAL APPROACH TO CALCULATE *B* FOR TOROIDAL CURRENT

The geometry of a current carrying toroidal conductor, having a uniform or a profiled current density, is completely characterized by its major radius ($R_0$), minor radius ($a_0$), elongation ($\kappa$) and triangularity ($\delta$) [1]. As the present study is restricted to the toroidal current with circular cross-section only, the last two parameters are respectively one and zero always and, an understanding of the complete geometry requires only $R_0$ and $a_0$, as given in Fig. 1. To calculate the theoretical value of magnetic field due to the entire toroidal current channel, carrying current $I$, the current column is assumed to be composed of $N$ number of co-axial filaments with different radii. The current is distributed according to the current density profile:

$$J(r) = J_0 \left(1 - \frac{r^2}{a_0^2}\right)^\gamma \qquad (1)$$

among the filaments, where $J_0$ is the peak value of $J$ at $r = 0$, $\gamma$ is the profile exponent and $r$ is the distance from the center of current channel 'C', as shown in Fig. 1. Finally, the resultant magnetic field ***B*** is calculated using the principle of superposition for magnetic field at any subject point due to all co-axial filaments, each of which carries ($I/N$) amount of current. The calculated values for ***B***, using this technique, is compared with those using the available magnetic field solver software, like, POISSON SUPERFISH, FEMM etc. and a good amount of agreement with a maximum uncertainty < 3% is achieved.

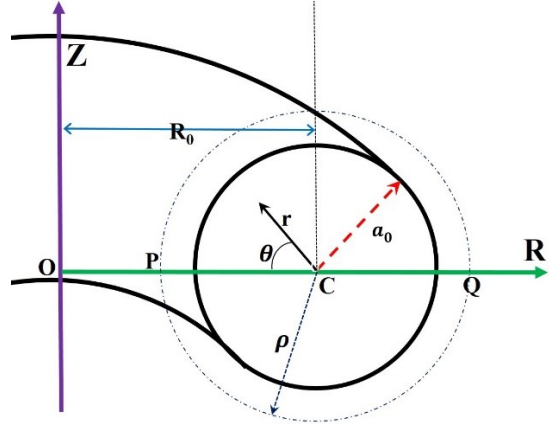

FIG 1. Poloidal cross-sectional view of the right half for a toroidal conductor with major radius $R_0$ and minor radius $a_0$.

### III. NATURE OF POLOIDAL MAGNETIC FIELD AROUND TOROIDAL CONDUCTOR

The filamentary approach is used to calculate **B** for a toroidal uniform current density, carrying total 1 kA of current, at P ($\theta = 0°$) and Q ($\theta = 180°$), both of which are situated on an arc of radius $\rho$, having center at C, as shown in Fig. 1. The geometry of this current channel is defined by its major radius $R_0 = 75$ cm and varying minor radii values, as enlisted in Table 1. For the simplicity of understanding, center of current column is always kept at 'C' and only $a_0$ is varied. The magnitude of **B** at P and Q for a particular value of current cross-section, given by $a_0$, ensures that **B** rises with a drop of $a_0$ at P, which is in contrary with the same at Q. A detailed analysis is addressed in reference [5-6] using POISSON/SUPERFISH and hence a point of invariance (PoI) is found in between P and Q on the arc of radius $\rho$ at an angle of $\theta = \theta_I(\rho)$, which is named as the angle of invariance (AoI). The entire analysis finally showed that $\theta_I$ follows a linear relation with ($\rho/R_0$), as given in equation 1, irrespective of the geometry of the toroidal current.

$$\theta_I = \left[1.297 - 0.571 \left(\frac{\rho}{R_0}\right)\right] \text{Radian} \qquad (2)$$

TABLE I. Magnetic field due to 1 kA of toroidal current with different circular cross-sectional area, i.e., minor radius ($a_0$), assuming a uniform current density. Major radius $R_0 = 75$ cm, $\rho = 27.5$ cm.

| Sl. No. | $\theta°$ | $a_0$ (cm) | B (G) |
|---|---|---|---|
| 1 | 0 | 25.0 | 12.5125 |
| 2 | 0 | 12.5 | 12.7298 |
| 3 | 0 | 6.5 | 12.8352 |
| 4 | 180 | 25.0 | 4.5558 |
| 5 | 180 | 12.5 | 3.5867 |
| 6 | 180 | 6.5 | 3.4558 |

The consequences say that at $\theta_I$, no variation of **B** is observed due to the change in $a_0$ and this is true for any toroidal geometry.

### IV. EXPERIMENTAL ARRANGEMENTS

To have the required information of |B| for a toroidal current, the number of magnetic probes should suffice to measure |B| around the current channel at a close vicinity at as many points as possible. The experimentally measured |B| due to the Ohmic plasma in Aditya Upgrade tokamak using a Mirnov probe garland [8-9], consisting of total 16 probes, is taken into account for the analysis of experimental data. Fig. 2 schematically shows the Mirnov

garland and its 16 probes with specific nomenclature. All these probes are installed on a single poloidal plane with an angular separation of 22.5° from each other. So, probe number 8 (M8), for example, is specified by $\theta = 11.25°$ also, and so on for other probes. According to the orientation, these magnetic probes are subjected to capture the total |B| due to the plasma current, along with all other unwanted magnetic fields, generated by surrounding currents. Firstly, the captured |B| from a single probe is corrected for all unwanted magnetic pick-ups using the data for vacuum shots, in which all the surrounding coils are driven with low parameters of current, though no plasma discharge appears. This correction essentially discards all the additional unwanted poloidal magnetic fields [9]. As Mirnov probes are subjected to capture magnetic signals from a wide range of frequencies [10], the present study is subjected to deal with the experimental data that are having a very low frequency. So, the data is further processed through a low pass filter below 30 Hz. Moreover, the analysis needs to measure the horizontal and vertical movement of plasma using COSINE and SINE coils respectively [9]. Finally, the acquired data is corrected using the laboratory calibration factors for each of the 16 probes to get the correct magnitude of magnetic field at probe locations.

## V. EXPERIMENTAL MEASUREMENT OF REQUIRED EXTERNAL MAGNETIC FIELD

The numerical results, discussed so far, deals with a plasma column that has its center at C, i.e., with constant $R_0$, though $a_0$ is varied and **B** is needed to be measured with these compliances to establish the experimental validation of those numerical observations. Though this condition is not very easy to achieve experimentally in case of toroidal plasma in a tokamak, a subtle way of achieving this condition is adapted using the experimental measurements. With the privilege of having 16 Mirnov probes, distributed on a single poloidal plane at an equal angular separation of 22.5° from each other, the expected measurement is achieved for plasma discharges using a data from a pair of Mirnov probes that are situated at $\pm\theta$ locations. Before discussing the experimental data analysis in further details, the technique is established numerically, as given in the following sub-section.

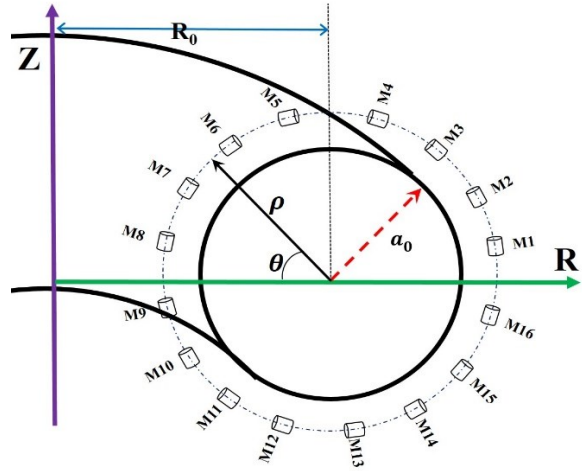

FIG 2. Mirnov probe assembly, installed in Aditya-U tokamak.

### A. Implementation of the idea of variation in cross-sectional area with a fixed center

The magnetic field |B| is required at, say, $\rho$ = d, $\theta$ = 11.25° (according to Fig. 2) for the plasma column that has a major radius $R_0$ and a minor radius $(a_0 - \epsilon)$ after a sustainable vertical movement of the column by $\pm\epsilon$ amount. Due to the prior condition of measuring |B| for plasma with its center at C and with a minor radius $(a_0 - \epsilon)$, the required |B| at M8 is achieved by algebraic averaging of |B| from a pair of probes that are situated at $\theta = \pm 11.25°$. To verify this algebraic averaging to work appropriately, |B| is calculated theoretically at different poloidal angles $(\theta)$ due plasma column, having uniform and peaked current densities, as given in table 2. Basically, |B| is calculated for toroidal plasma that has major and minor radii of 75 cm and 22.5 cm respectively; but, in first case current center being on Z = 0 plane and then it getting vertically shifted by 2.5 cm. The corresponding estimated |B| for vertically shifted current from C using the averaged magnetic fields from two vertically symmetric (i.e., $\pm\theta$) probes remains the same with the |B| at either probe for plasma remaining at C, though minor radius becoming $(a_0 - \epsilon)$, and it introduces a maximum uncertainty of 0.9%, irrespective of current density profiling. With this idea, the experimental data are collected and analysed, as given in the following sub-section.

TABLE II. Averaged magnetic field at different poloidal locations due to 1 kA of toroidal current with circular cross-section (major and minor radii being 75 cm and 22.5 cm respectively).

| Probe pair | dZ (cm) | Average total B (G) | |
|---|---|---|---|
| | | $\gamma = 0$ | $\gamma = 5$ |
| <8,9> | 0 | 12.2776 | 12.3608 |
| <8,9> | 2.5 | 12.2308 | 12.3056 |
| <7,10> | 0 | 11.2641 | 11.282 |
| <7,10> | 2.5 | 11.2399 | 11.2769 |
| <6,11> | 0 | 9.6375 | 9.5995 |
| <6,11> | 2.5 | 9.7067 | 9.6479 |
| <5,12> | 0 | 7.8984 | 7.8299 |
| <5,12> | 2.5 | 7.9439 | 7.8968 |
| <4,13> | 0 | 6.3649 | 6.2693 |
| <4,13> | 2.5 | 6.4199 | 6.3186 |
| <3,14> | 0 | 5.1428 | 5.0284 |
| <3,14> | 2.5 | 5.1715 | 5.0476 |
| <2,15> | 0 | 4.2893 | 4.1516 |
| <2,15> | 2.5 | 4.2756 | 4.1497 |
| <1,16> | 0 | 3.9835 | 3.692 |
| <1,16> | 2.5 | 3.9895 | 3.6807 |

**B. Experimental results**

For the experimental validation of the variation in |B| due to the change in cross-sectional area of the current channel, a number of plasma discharges of Aditya Upgrade tokamak are analysed. The entire analysis of experimental data is based on the empirical fact that the cross-section of toroidal plasma column remains circular even after a sustainable amount of movement along any direction (horizontal or vertical). The judicial choice of time for picking the measured |B| due to the plasma current is made on the basis of the outcomes from previous sub-section and at zero horizontal movement (i.e., $dX = 0$), though having different vertical shifts ($dY$). When the plasma column is shifted in vertical direction only by $\epsilon$ amount (i.e., $dY = \epsilon$), keeping the horizontal displacement to be approximately zero (i.e., $dX \sim 0$), the averaged magnetic field from a symmetric pair of magnetic probes gives a measurement of |B| for the plasma with reduced minor radius $(a_0 - \epsilon)$ and having centre coinciding with the geometric center C. For example, times for considering |B| from all magnetic probes are chosen when $dX \sim 0$ cm, as measured by COSINE coil [9]. Table 3 summarizes the measured |B| from all the 8 magnetic probe pairs for two typical plasma discharges at selected times. As the Mirnov probes are radially 27.5 cm away from the geometric center of plasma, i.e., C, $\theta_I$ comes up to be $\theta_I \approx 62.32°$ for $R_0 = 75$ cm, according to equation 1. Due to the rise in $a_0$ of the plasma, the magnetic field |B| shows a drop at all three probes, for which $\theta < \theta_I$, for all plasma discharges, analysed so far and it is evident from Table 3. In contrary, |B| increases with the rise of $a_0$ at all 5 probe locations from 78.75° to 168.75°. Thus, $\theta_I$ must lie between 56.25° and 78.75°, which is suggested by equation 1. By looking into the data for fifty typical plasma discharges, the value of $\theta_I$ is further confirmed to remain in between 56.25° and 78.75°, in agreement with 62.32° for this given geometry.

**VI. DISCUSSION**

The variation of magnetic field |B| is calculated as well as measured for 1kA current throughout this paper for a fair comparison. As the toroidal fusion machines deal with a current of MA order of magnitude, the variation in |B| becomes very much significant in reality. For instance, according to Table-3, the change in B due to shrinking of the minor radius turns out to be tens order of magnitude in Gauss, even in case of Aditya-U like medium sized tokamak, in which plasma current is generated of the order of few hundreds of kilo Amperes. Thus, such a change in |B| needs

to be carefully measured to understand the complete topology of |B| during a change in minor radius of MA order of plasma current, especially where a preciseness of |B| is required. Here, it is to be emphasized that the order of error, which is introduced due to the averaging of |B| from two symmetric magnetic probe pair, is much less than the change of |B|, occurring due to the change in $a_0$. This is evident from a fair comparison between Table-2 and 3.

When the current density has a peaked profile, rather a uniform one, the effective current remains in the central region of the column and the entire column behaves as a squeezed plasma only, keeping all the outcomes unaltered from the uniform one. In case of a peaked profiling of current density, as seen from Table-2, a maximum change in |B| due to a change in the cross-section of the current column appears to be (~7.0/782.99% =)0.9%, whereas the same is 0.05% for uniform current distribution. Thus, the inferences of this analysis is not affected by the change in the current density profile. This fact is also supported by the Ampere's law, which predicts that the external magnetic field |B| due to the entire current column does not depend on the current density pattern, rather on the total enclosed current [11-12].

If a Grad-Shafronauv shift of the plasma centroid towards outboard by ~ 2.5 cm is considered for Aditya-U plasma [13], no significant change in the outcomes of the analysis is expected, except in the value of $\theta_I$ changing from 62.3° to 62.7°. So, this shift does not impact the consequences in an adequate manner.

TABLE III. Experimental averaged magnetic field at different poloidal locations due to 1 kA of toroidal current with circular cross-section (major radius being 75 cm).

| Shot | Time (ms) | dX (cm) | dY (cm) | $a_0$ (mm) | B (G) for probe pair | | | | | | | |
|---|---|---|---|---|---|---|---|---|---|---|---|---|
| | | | | | <8,9> (11.25°) | <7,10> (33.75°) | <6,11> (56.25°) | <5,12> (78.75°) | <4,13> (101.25°) | <3,14> (123.75°) | <2,15> (146.25°) | <1,16> (168.75°) |
| 38943 | 51.24 | 0.01 | -0.35 | 246.5 | 3.9004 | 2.6517 | 2.6729 | 2.061 | 1.1359 | 1.1154 | 0.8091 | 0.1828 |
| 38943 | 80.09 | 0.01 | 0.13 | 248.7 | 1.7136 | 1.4522 | 1.9334 | 2.13 | 1.4785 | 2.0561 | 1.9301 | 0.8529 |
| 37016 | 20.53 | 0.00 | -0.81 | 241.9 | 4.2458 | 3.374 | 3.2243 | 2.3932 | 1.4252 | 1.4257 | 0.9452 | 3.17E-04 |
| 37016 | 78.82 | 0.00 | -0.05 | 249.5 | 1.5607 | 1.4286 | 3.0769 | 2.5575 | 2.5967 | 2.4751 | 2.4343 | 9.99E-04 |

## VII. CONCLUSIONS

The change in the cross-sectional area of a toroidal current with circular cross-section has a strong impact on its surrounding external magnetic field ***B*** and the topology of ***B*** is severely impacted by the overall cross-sectional geometry of the torus. This numerical observation is experimentally verified in this work. Firstly, the change in $a_0$ without a change in $R_0$ is successfully addressed in the experimental scenario and hence the change in ***B*** at specific probe locations is found. The empirical data shows that the change of ***B*** due to shrinking of minor radius has an impact on ***B*** at inboard and this is completely opposite to those at outboard, which was earlier predicted by numerical results [5]. Thus, an angular location $\theta_I$ from which the nature of change in ***B*** appears, is empirically found to lie in between 56.25° and 78.75°. Theoretically this $\theta_I$ is calculated to be 62.32°, which agrees with the experimental observation well.

## ACKNOWLEDGMENTS

The authors acknowledge the electronics team, power supply team as well as the data acquisition team for their supports during the execution of experiment and acquisitions.

**References:**

[1] J. Wesson, *Tokamaks* (Oxford:Clarendon, 1997).
[2] Y. Shimomura et al., Nucl. Fusion 39, 1295 (1999).
[3] V. Mukhovatov et al., Plasma Phys. Control. Fusion 45, A235 (2003).
[4] L. L. Lao et al., Nucl. Fusion 30 (6), 1035 (1990).
[4] L.L. Lao et al., Nucl. Fusion 25, 1611 (1985).
[5] S. Aich, J. Thakkar, and J. Ghosh, Plasma Fusion Res. 17, 2403055 (2022).
[6] S. Aich, A. Iyer, and J. Ghosh, IEEE Transactions on Plasma Science 52 (7), 2492-2499 (2024).
[7] Reference Manual for the Poisson/Superfish Group of Codes, document LA-UR-87-126, Los Alamos Accelerator Code Group, Los Alamos National Laboratory, New Mexico, 1987.
[8] S.V. Mirnov, Soviet Atomic Physics 30, 22 (1971).
[9] S. Aich et al., Plasma Res. Express 3, 035005 (2021).
[10] G. Hammett and K. McGuire, *Analysis of Mirnov Oscillations on PDX* (PPPL, Princeton, 1982).
[11] D. J. Griffiths, *Introduction To Electrodynamics* (USA: Prentice-Hall, 1981).
[12] J. D. Jackson, *Classical Electrodynamics* (USA: Wiley, 1962).
[13] R. L. Tanna et al., Nucl. Fusion 57 (10), 102008 (2017).